\documentclass[11pt]{article}
\usepackage{amsmath}
\usepackage{amssymb}
\usepackage{graphicx}
\numberwithin{equation}{section}
\begin{document}
\setcounter{page}{0}
\thispagestyle{empty}
\begin{flushright}
\end{flushright}
\vspace*{2.0cm}
\begin{center}
{\large \bf Fine Structure Constant, Domain Walls and  Generalized 
\\
Uncertainty  Principle in the Universe}
\end{center}
\vspace*{1cm}
\renewcommand{\thefootnote}{\fnsymbol{footnote}}
\begin{center}
{%
Luigi Tedesco$^{1,2}$\protect\footnote{Electronic address: {\tt
luigi.tedesco@ba.infn.it}} \\[0.5cm]  $^1${\em
Dipartimento di Fisica, Universit\`a di Bari, I-70126 Bari, Italy}
\\[0.2cm] $^2${\em INFN - Sezione di Bari, I-70126 Bari, Italy}}
\end{center}

%
%
\renewcommand{\abstractname}{\normalsize Abstract}
%
%
\vskip 1.5truecm
\begin{abstract}
In this paper we study the the corrections to the fine structure
constant from the generalized uncertainty principle in the spacetime
of a domain wall. We also calculate the corrections to the standard
formula to the energy of the electron in the hydrogen atom  to the
ground state, in the case of spacetime of a domain wall and
generalized uncertainty principle. The results generalize the cases
known in literature.
\end{abstract}
%
%
%
%
\newpage
%
%
%
%
\renewcommand{\thesection}{\normalsize{\arabic{section}.}}
\section{\normalsize{Introduction}}
\renewcommand{\thesection}{\arabic{section}}
In the last years there has been an interest in cosmology with a space-time variation of the 
constants of nature.
In the 1920 in order to explain the relativistic splits of the
atomic spectral lines, Arnold Sommerfeld introduced the fine structure
constant
\begin{equation}
\label{alpha1}
\alpha_0=\frac {e^2} {4 \, \pi \epsilon_0 \, \hbar \,c}
\end{equation}
where $c$ is the speed of light in vacuum, $\hbar=h/2 \pi$ is the
reduced Planck constant, $e$ is the electron charge magnitude and
$\epsilon_0$ is the permettivity of free space, all quantities
measured in the laboratories on Earth.  The numerical value of the
constant is $\alpha_0 \sim 1/137.035999710$ \cite{uzan} that can be
determined without any reference to a specific system of the units
and $\alpha$ gives the strength of the electromagnetic interaction.
In the recent years  possible variations of the fine structure
constant have been observed, these observations suggest that about
$10^{10}$ years ago $\alpha$ was smaller than today. On the other
hand time variation of fundamental constants has been an intriguing
field of theoretical research since the propose by Dirac in 1937
\cite{dirac} where in the large numbers hypothesis he conjectured
that the fundamental constants are functions of the epoch. The
physical motivation to search a time or a space dependence of
fundamental constants originates because the effort to unify the
fundamental constant imply variations of the coupling constants
\cite{barrow}. 
Let us introduce $\alpha(z)$ that is the value that might be dependent on the time.
The variations of $\alpha$  can be measured by the so
called "time shift density parameter"
\begin{equation}
\label{alpha2}
\frac {\Delta \alpha} {\alpha} \equiv \frac {\alpha(z) - \alpha_0}
{\alpha_0}
\end{equation}
with $\alpha_{0}$ value of $\alpha$ today.
\\
From an experimental point of view there are two ways to test the
validity of the "constant" hypothesis of $\alpha$: local and
astronomical methods. The former connected with local geophysical
data, the natural reactor $1.8 \times 10^{9}$ years ago ($z \sim
0.16$) in Oklo \cite{oklo}, these data give \cite{petrov} $|
\dot{\alpha} / \alpha| = (0.4 \pm 0.5) \times 10^{-17}\, yr^{-1}$
(or $|\Delta \alpha/\alpha| \leq 2 \times 10^{-8}$) that is one of
the most stringent constrain on the variation of $\alpha$ over
cosmological time scales. The latter methods consider deep-space
astronomical observations, they mainly consider the analysis of
spectra from high red-shift quasar absorption system. Evidence of
time variation of $\alpha$ derive from these data \cite{webb}. It is
important to say that these data, coming from the Keck telescope in
the Northerm hemisphere, give for a range of the red-shift $0.2 < z
< 4.2$ \cite{murphy}: $ \Delta \alpha / \alpha = (-0.543 \pm 0.116)
\times 10^{-5}$. If we assume a linear increase of $\alpha$ with
time, we have a drift rate  $d \,\text{ln} \alpha/ dt = (6.40 \pm
1.35) \times 10^{-16}$ per year. In any case $\Delta \alpha /
\alpha$ may be more complex \cite{marciano} and a linear
extrapolation may not be valid when we consider a cosmic time scale.
However, independent analysis of the same phenomena with VLT
telescope, in Chile, does not find any variation of $\alpha$
\cite{srianand}, in fact we find $\Delta \alpha / \alpha = (-0.06
\pm 0.06) \times 10^{-5}$. There is an intensive debate in
literature about possible reasons for disagreement, for example a
possible reason may be that the Keck telescope is in the Northerm
hemisphere and VLT telescope is in the Southerm hemisphere. Recently
\cite{murphy2}  a re-analysis of Ref. \cite{srianand} varying
$\alpha$ by means of the multiple heavy element transition on the
Southern hemisphere has been reported, obtaining $\Delta \alpha /
\alpha = (- 0.64 \pm 0.36) \times 10^{-5}$. On the other hand this search may connected in astronomical observations for variations
in the fundamental constants in quasar absorption spectra and in laboratory \cite{berengut}. 
\\
The experimental physics  has reached very high precision therefore in order 
to search  corrections very fine to our theories  in the description 
of the nature it is necessary to introduce logical systems more and more sophisticated. In this
context, to search corrections to the   fine structure constant, it is only possible if we study very complex fields of the knowledge. 
The conceptual utilization of the GUP may be usefull in order to calculated the corrections to the fine structure constant. 
The paper follows this line in which we want to built a bridge between corrections to the alpha and GUP. On the other hand 
if we consider a cosmological ambit these corrections may have important  consequences if we also consider
a topological defect has a domain wall on large scale in the universe. 
For these reasons it is important to employ gup and $\alpha$ evolution. 
\\
\\
The search for a quantum theory of the gravitation is one of the
most intriguing problems in physics. The generalized uncertainty principle is a consequence
of incorporating a minimal lenght from a theory of quantum gravity. When we consider a quantum gravity theory 
we need a fundamental distance scale of order the Planck lenght $l_p$. These reasonings  induce the possibility 
to have corrections to the Heinsenberg principle in order to have a more general uncertainty principle (GUP). Thus the 
Heinsenberg principle
\begin{equation}
\label{eq1}
\Delta x \, \Delta p \gtrsim \hbar
\end{equation}
has to be replaced by
\begin{equation}
\label{eq2}
\Delta x \, \Delta p \gtrsim \hbar + \beta \, l^2_{p} \, \frac
{{\Delta p}^2} {\hbar}
\end{equation}
here $\Delta x$ and $\Delta p$ are the position and momentum uncertainty for a quantum particle,
$\beta$ is a positive dimensionless coefficient that may
depend on the position $x$ and momentum $p$, usually assumed to be
of order one and $l_p = (G \hbar /c^3)^{1/2} \sim 1.66 \times
10^{-33}$ cm is the Planck length. 
It is important to stress that $l^2_p {\hbar}$ may be replaced with the Newtonian 
constant G, therefore the second term in (\ref{eq2})  is a consequence of gravity.
The physical reason consider that
the quantum mechanics limits the accuracy of the position and
momentum of the particle by the well known rule $ \Delta x \geq
\hbar/ \Delta p$, moreover if we consider general relativity the
energy cannot be localized in a region smaller than the one defined
by its gravitational radius, $\Delta x \geq l^2_{Pl} \, \Delta p$.
If we combine the results, there is a minimum observable length
$\Delta x \geq \text{max}( 1/ \Delta p; l^2_{p} \, \Delta p) \geq
l_{p}$. This final result is the Generalized Uncertainty Principle,
that can be summarized as eq.(\ref{eq2}).
\\
Generally speaking the GUP is obtained when the Heinsenberg
uncertainty principle is considered combining both quantum theory
and gravity and it may be obtained from different fields and
frameworks as strings \cite {amati}, black holes \cite {maggiore}
and gravitation \cite {scardigli}, where the gravitational
interaction between the photon and the particle modifies the
Heinsenberg principle, adding an additional term in eq. (\ref{eq2})
proportional to the square of the Planck length $l_{p}$. From a
physical point of view very interesting consequences can be found in
\cite{Brau}.
\\
\\
The initial  stages of the primordial Universe according to the standard model of the particles physics, are
often described as the era of the phase transition.
In the recent years the cosmological consequence of primordial phase
transitions  has been the subject of many studies in the early
Universe.  When we have a cosmological phase transition, topological
defects necessarily can be formed \cite{{kibble,vilenkin}}: they are domain walls, cosmic strings or monopoles. These phenomena are expected to be produced at a phase transition in
various area of physics, for example also in condensed matter
physics several examples have been observed, while up today in
particle physics, astrophysics and cosmology it is not the case; on the other hand they could have very important
cosmological consequences.
Generally people study cosmic strings because they present
interesting properties and there are not any bad cosmological
consequences, instead domain walls scenarios have attracted less
attention since there is the so called Zeldovich bound
\cite{zeldovich}, in which in a linear scaling regime would dominate
the energy density of the Universe violating the observed isotropy
and homogeneity. A domain wall network was proposed to explain dark
matter and dark energy \cite{spergel}.
\\
The connection between topological defects and variation of the
fundamental constants is an intriguing field of work. The
corrections to the fine structure constant has been calculated in
the spacetime of a cosmic strings \cite{nasseri1}. In a recent paper
\cite{campanelli} it has been studied the correlation of time
variation of the fine structure constant in the space time of a
domain wall and in particular it has been shown that the
gravitational field generated by a domain wall acts as a medium with
spacetime dependent permettivity $\epsilon$. In this way the fine
structure constant will depend on a time-dependent function at a
fixed point. A further step has been obtained with the calculation
of the corrections to the fine structure constant in the spacetime
of a cosmic string from the generalized uncertainty principle
\cite{nasseri2}. In this paper we study the corrections to the fine
structure constant in the spacetime of a cosmic domain wall taking
into account the generalized uncertainty principle, are calculalted. In other terms we generalize our previous
study \cite{campanelli}. The paper is organized as follows: in Section 2 we summarize our previous results
obtained considering the time variation of the fine structure constant in the space time of a cosmic domain wall, in Section 3 we generalized the results taking into account the generalize uncertainty principle, in Section 4 we calculate, as application, the correction to the energy ground state of the hydrogen atom, the results are summarized in the concluding Section 5.
\renewcommand{\thesection}{\normalsize{\arabic{section}.}}
\section{\normalsize{$\alpha$ in the spacetime of a domain wall}}
\renewcommand{\thesection}{\arabic{section}}
As it is well known a domain wall is a topologically stable kink
produced when a vacuum manifold of a spontaneously broken gauge
theory is disconnected \cite{vilenkin}. A very important concept
regards the surface energy density $\sigma$ of a domain wall because
it determines the dynamics and gravitational properties, but
unfortunately $\sigma$ is very large and this implies that cosmic
domain walls would have an enormous impact on the homogeneity of the
Universe. It is possible to have constraint on the wall tension
$\sigma$ from the isotropy of the cosmic microwave background, in
fact if  a few walls stretch across the present horizon we have an
anisotropy fluctuation temperature of CMB $\frac {\delta T} {T} \sim
2 \pi G \sigma H_O^{-1}$ with G Newton's constant and $H_0$ Hubble
constant. The anisotropy $\frac {\delta T} {T} \leq 3 \times
10^{-5}$ arises from WMAP therefore it is not possible to have
topologically stable cosmic walls with $\sigma \geq 1 Mev^3$.
\\
A cosmic domain wall in the Universe modifies the electromagnetic
properties of the free space and in particular if we consider the
gravitational field generated by a wall, it acts as a medium with
space and time dependent permettivity. Therefore eq. (\ref{alpha1})
implies that the fine structure constant at fixed point will be a
time-dependent function. In this Section we follow the way of \cite{campanelli}.
\\
Let us consider the line element associated to the spacetime of a
thin wall \cite{vilenkinPLB}
\begin{equation}
\label{metric}
ds^2= e^{-4 \pi G \sigma \mid x \mid}\,(c^2 \, dt^2-dx^2) - e^{4 \pi
G \sigma (ct-\mid x \mid)} \, (dy^2+dz^2),
\end{equation}
in which we have considered a model with infinitely static domain
walls in the $zy$-plane.
Generally speaking in a curved spacetime the electromagnetic field
tensor $F_{\mu \nu}$ has electric and magnetic fields respectively
defined as
\begin{equation}
\label{EB}
E_i= F_{0\, i} \,\,\,\,\,\,\,\,\,\, B^i= - \frac {1} {2
\sqrt{\gamma}} \, \epsilon^{ij k} \, F_{jk},
\end{equation}
with $\gamma= \text{det} \parallel \gamma_{ij} \parallel$
determinant of the spatial metric and $\epsilon^{ij k}$ Levi-Civita
symbol. If we consider a charged particle $q$, the charge density at
rest in ${\bf x} = {\bf x_0}$ is
\begin{equation}
\label{density}
\rho= \frac {q} {\sqrt{\gamma}} \delta ({\bf{x}}-{\bf x_0}).
\end{equation}
We write the divergence and curl operators in curved spacetime as
\begin{equation}
\label{div}
\text{div}\,  {\bf v}= \frac {\partial_i (\sqrt{\gamma}\,  v^i)}
{\sqrt{\gamma}}
\end{equation}
and
\begin{equation}
\label{curl}
(\text{curl} \, {\bf v})^i = \frac { \epsilon^{ijk} (\partial_j v_k
-
\partial_k v_j)} {2 \sqrt{\gamma}}
\end{equation}
respectively, therefore Maxwell's equation in three dimensions are
\begin{equation}
\label{BE}
\text{div} \,{\bf B} = 0 \,\,\,\,\,\,\,\,\,\,\,\,\,\,\,\,
\text{curl}\, {\bf E} = - \frac {1} { \sqrt{\gamma}} \frac {\partial
(\sqrt{ \gamma}\, {\bf B})} {\partial t},
\end{equation}
\begin{equation}
\label{DH}
\text{div} \,{\bf D} = 4 \pi \rho \,\,\,\,\,\,\,\,\, \text{curl}\,
{\bf H} =  \frac {1} { \sqrt{\gamma}} \frac {\partial (\sqrt{
\gamma}\, {\bf D})} {\partial t}.
\end{equation}
where 
\begin{equation}
\label{DH2}
\text{\bf D}= \frac {{\text {\bf E}}} {\sqrt{g_{00}}}  \;\;\;\;\;\;\;
\text{\bf H} = \sqrt{g_{00}} \text{\bf B} \, .
\end{equation}
If we indicate with $\nabla$ the three-dimensional nabla operator in
Euclidean space, we can rewrite the first equation of eq. (\ref{DH})
as
\begin{equation}
\label{eE}
\nabla \cdot (\epsilon \, {\bf E}) = 4 \pi q \delta ({\bf x} - {\bf
x_0}),
\end{equation}
where $\epsilon=\sqrt{\gamma}/ \sqrt{g_{00}}.$ The solution of
Poisson equation, eq.(\ref{eE}), is  $ \epsilon \, {\bf E} = q /{ 4
\pi \epsilon \, r^3}$ that gives for the electric field the
expression
\begin{equation}
\label{E}
{\bf E} = \frac {q} {4 \pi \epsilon \, r^3} \, {\bf r}.
\end{equation}
It is interesting to note that if we consider the metric
(\ref{metric}), a domain wall produce a gravitational field that
acts as a medium with a permettivity $\epsilon$ that has the
expression
\begin{equation}
\label{epsilon}
\epsilon= \epsilon_0 \, e^{ 4 \pi G \sigma (c t - |x|)}.
\end{equation}
Therefore  a cosmic domain wall in the Universe modifies the
electromagnetic properties of the free space  and  taking into
account eq.(\ref{epsilon}), we can say that in the free space the
constant $\alpha$ is  given by eq.(\ref{alpha1}) and in the
spacetime of a domain wall is
\begin{equation}
\label{newalpha}
\alpha=\frac {e^2} {4 \, \pi \epsilon \, \hbar \,c}\,.
\end{equation}
that is to say the fine structure constant in the spacetime of a
domain wall is spacetime dependent.
%
%
%

%
%
\vskip 2truecm
\renewcommand{\thesection}{\normalsize{\arabic{section}.}}
\section{\normalsize{$\alpha$ in the spacetime of a domain wall from the
generalized uncertainty princicle}}
\renewcommand{\thesection}{\arabic{section}}
\vskip 0.6truecm
Now we calculate the corrections to the fine structure constant in
the spacetime of a domain wall taking into account the generalized
uncertainty principle. If we take into account the gravitational
interactions, the Heinsenberg principle must be revised with the
generalized uncertainty principle, that is to say  $\Delta x \,
\Delta p \geq \hbar$ becomes $\Delta x \, \Delta p \gtrsim \hbar +
\beta \, l^2_{Pl} \, ({{\Delta p}} / {\hbar})^2$, this suggests to
introduce a kind of "effective" Planck constant, $h_{eff}$, due to
the generalized uncertainty principle, defined as
\begin{equation}
\label{heff}
{\hbar}_{eff} = \hbar \left[ 1 + \beta \, l^2_{Pl} \left ( \frac
{\Delta p} {\hbar} \right)^2 \right]
\end{equation}
in order to write $\Delta x \, \Delta p \geq {\hbar}_{eff}$.
Therefore the constant will be
\begin{equation}
\label{alphadwgup}
\alpha_{eff}= \frac {e^2} {4 \, \pi \,  \epsilon \, \hbar_{eff}}
\end{equation}
with $\epsilon$ given by (\ref{epsilon}). In this way the GUP is be able to introduce "itself"
in the expression and  change the structure of $\alpha$.
\\
In order to obtain
$\alpha_{eff}$ let us consider eq. (\ref{eq2}) that we solve as a
second order equation for the momentum uncertainty in terms of the
distance uncertainty, we have
\begin{equation}
\label{deltap1}
\frac {\Delta p} {\hbar} = \frac  {\Delta x} {2 \,  \beta \,
l^2_{Pl}} \left[1 - \sqrt {1 - \frac {4 \, \beta \, l^2_{Pl}}
{(\Delta x)^2}} \right]
\end{equation}
(we do not consider the sign + in the parenthesis because non
physical, in fact if we impose correct classical limit $l_{pl} \rightarrow 0$ we only have minus sign).
\\
We obtain $\Delta x$ considering  Bohr's radius in the spacetime of
a domain wall. In absence of a domain wall a Bohr's atom has the
radius (n=1) $r_0= 4 \pi \epsilon_0 {\hbar}^2 / m e^2$, with $m$
mass of the electron,  but in presence of a domain wall and the GUP,
it becomes
\begin{equation}
\label{radius}
\tilde{r}_{0}= \frac {4 \, \pi \, \epsilon \, {\hbar}^2} { m \, e^2}
\equiv \Delta x.
\end{equation}
In other terms Bohr's radius in  a spacetime of a domain wall,
$\tilde{r}_0$, is connected with  $r_0$ classical Bohr's radius by
the relation
\begin{equation}
\label{radius2}
\tilde{r}_0 = r_0 \, e^{4 \pi G \sigma (c t-|x|)}.
\end{equation}
Now introducing (\ref{deltap1}) in (\ref{heff}) we obtain $h_{eff}$
as a function of $\Delta x$. This $h_{eff}$ introduced in
(\ref{alphadwgup}), finally gives the fine structure constant in the
spacetime of a domain wall with  the generalized uncertainty
principle:
\begin{equation}
\label{alphafinale}
\alpha_{eff} = \frac {e^2} {4 \pi \epsilon \, c \, \hbar} \left[ 1 +
\frac {(\Delta x)^2} {4 \, \beta \, l^2_{Pl}} \left(1- \sqrt{1-
\frac {4 \, \beta \, l^2_{Pl}} {(\Delta x)^2}}\right)^2
\right]^{-1}.
\end{equation}
We discuss eq. (\ref{alphafinale}) starting from the case without
the spacetime of a domain wall, in other terms $\alpha$  with the
generalized uncertainty principle. There are several studies
\cite{Nozari} that consider  non-commutativity spacetime and quantum
gravitational effects in the calculation of the fine structure
constant
with $\Delta x$ given by (\ref{radius}). If we  only consider the
GUP effect on the fine structure constant we have
\begin{equation}
\label{alphagup}
\alpha_{gup} \simeq \alpha_0 \, [1- 3.6 \times 10^{-50}]\, ,
\end{equation}
but in presence of the cosmic domain wall it is possible to render
explicit the expression of $\alpha$
\begin{equation}
\label{alphaesplicito}
\alpha_{eff} = \alpha_0 \, e^{- 4 \pi G \sigma (ct - |x|)} \left[1+
\frac {r_0^2} {4 l^2_{Pl}} e^{8 \pi G \sigma (ct-|x|)} \left( 1-
\sqrt{1- \frac {4 l^2_{Pl}} { r_0^2} e^{-8 \pi G \sigma (ct - |x|)}}
\right)^2 \right]^{-1}
\end{equation}
%
%
\vskip 1.5truecm
\renewcommand{\thesection}{\normalsize{\arabic{section}.}}
\section{\normalsize{Corrections to the energy groung state of hydrogen atom}}
\renewcommand{\thesection}{\arabic{section}}
It is interesting to calculate the corrections to the energy  ground
state $E_0$ of the hydrogen atom in presence of a domain wall and
considering the GUP. Classically the hydrogen atom decays and it is
just the Heinsenberg Uncertainty Principle that assures the
stability. The energy of the electron in the hydrogen atom is
\begin{equation}
\label{classicalenergy}
E_{gup}^{dw} \sim \frac {p^2} {2m} - \frac {e^2} {4 \pi \epsilon \,
\tilde{r}_0},
\end{equation}
the GUP gives
\begin{equation}
\label{deltap}
\Delta p \geq \frac {\hbar} {\Delta x} + \frac {l^2_{Pl} \, (\Delta
p)^2} {\Delta x \, \hbar}.
\end{equation}
Now, let us iterate eq.(\ref{deltap}), neglecting the terms ${O}
(l^2_{Pl})$ and squaring both members we have
\begin{equation}
\label{p}
p^2 \geq (\Delta p)^2 \geq \frac {\hbar^2} {(\Delta x)^2} + 2 \,
\frac {{\hbar^2} \, l^2_{Pl}} {(\Delta x)^4}.
\end{equation}
Therefore eq. (\ref{classicalenergy}) for the energy becomes
\begin{equation}
\label{ene}
E_{gup}^{dw}= \frac {\hbar^2} {2 \, m \, {\tilde{r}_0}^2} - \frac
{e^2} {4 \, \pi \, \epsilon \, \tilde{r}_0} + \frac {\hbar^2 \,
l_{Pl}^2} {m \, \tilde{r}_0^4}.
\end{equation}
From a physical point of view, eq. (\ref{ene}) is very
interesting. If we "switch off" the domain wall contribution, the
first two terms on the second member, are the energy of the ground
state of the electron in the hydrogen atom, $E_0= - m\, e^2/ 8 \pi^2
\epsilon_0^2 \hbar^2= 13.6 \, \text{eV}$. The third term is the
correction to the ground state energy due to the generalized
uncertainty principle, that is to say:
\begin{equation}
\label{correction}
\Delta E_{gup} = \frac {m^3  \, l^2_{Pl} \, e^8} {(4 \pi
\epsilon_0)^4 \, \hbar^6} \sim 10^{-48} \text{eV}.
\end{equation}
This corrective term, due to the GUP, is very little to be
experimentally tested actually.
If now we "switch on" the domain wall contribution, we have
\begin{equation}
\label{Etotal}
E= - \frac {m \, e^4} {8 \, \pi^2 \, \epsilon_0 \,\hbar^2}\,  e^{- 8
\pi G \sigma (ct - |x|)} + \frac {m^3 \, e^8 \, l^2_{Pl}} {(4 \, \pi
\, \epsilon)^4 \, \hbar^6}\, e^{- 16 \pi G \sigma (ct - |x|)}.
\end{equation}
In other terms, when we consider the domain wall, the classical and
the GUP contribution to the energy are exponentially modulated,
therefore an integrate effect, starting from the early Universe, may
be relevant into the amplification to the correction to the energy
of the electron in a hydrogen atom from an experimental point of
view.
\renewcommand{\thesection}{\normalsize{\arabic{section}.}}
\section{\normalsize{Conclusion}}
\renewcommand{\thesection}{\arabic{section}}

In conclusion, if we consider that the gravitational interactions
may modify the Heinsenberg principle with the so called generalized
uncertainty principle and if we also consider that the fine
structure constant may be different in different epochs, it is
possible to study the right expression of the fine structure
constant in the spacetime of a domain wall, taking into account the
generalized uncertainty principle. In this paper we have examined
the effects of these two contributions on $\alpha$. We have found
the most general expression given by (\ref{alphaesplicito}).  The
modification of $\alpha$ involves two aspects, the domain wall's
contribution influences the value of $\epsilon_0$ that becomes
$\epsilon$ given by (\ref{epsilon}), while the GUP's contribution
acts in order to modify the Planck constant $\hbar$ into
$\hbar_{eff}$ given by (\ref{heff}).  $\alpha$ is very near at
$\alpha_0$ as we can see in (\ref{alphagup}), this means that the
GUP does not change the numerical  value in an appreciable way. The
domain wall's contribution consists into exponentially modulate the
$\alpha_0$ value and from a numerical point of view if we set
$ct-|x|= H_0^{-1}$,  we does not change the value of $\alpha$. On
the other hand it is possible to think as a kind of "integrate
effect" in the spacetime, in this way it is possible to have a
different evolution of $\alpha$ in the spacetime. 
These arguments are also very interesting because recently a sample of 153 measurements from the ESO Very Large Telescope indicate that $\alpha$ appears on
average to be larger than in the past \cite{webbking}. Moreover manifestations of a spatial variation in 
$\alpha$ must be independently confirmed by means terrestrial measurements as laboratory, meteorite data and nuclear reactor \cite{Berengut:2010ht} and by means a new test connected by big bang nucleosynthesis \cite{Berengut:2010yu}.
For completeness
we have also studied the corrections to the energy of the idrogen
atom if we  add both the actions: gup and domain wall. Also in this
case the corrections are still too small for the actual experiments.
Future investigations are in progress by the author.
\end{document}